\documentclass[lettersize,journal]{IEEEtran}

\usepackage{graphicx} 
\usepackage{amsmath}
\usepackage{float}
\usepackage{subfig}
\usepackage{caption}

\usepackage[boxed,ruled,vlined,linesnumbered,commentsnumbered, noresetcount]{algorithm2e}
\usepackage{algorithmic}

\usepackage{diagbox}
\usepackage{tikz}
\usetikzlibrary{shapes.geometric}
\usetikzlibrary{positioning}
\usepackage{soul,xcolor}
\setstcolor{red}

\usepackage{tabularx}
\usepackage{ragged2e} 
\newcolumntype{L}{>{\RaggedRight}X} 
\usepackage{lipsum} 
\usepackage{amsmath}
\usepackage{amssymb}

\makeatletter
\newcommand\notsotiny{\@setfontsize\notsotiny\@vipt\@viipt}
\makeatother

\begin{document}

\title{Towards Industry 5.0: Intelligent Reflecting Surface~(IRS) in Smart Manufacturing}

\author{Md. Noor-A-Rahim, Fadhil Firyaguna, Jobish John,   M. Omar Khyam,  Dirk Pesch~\IEEEmembership{Senior Member, IEEE,}  Eddie~Armstrong, Holger~Claussen,~\IEEEmembership{Fellow, IEEE,} and~H.~Vincent~Poor,~\IEEEmembership{Fellow, IEEE}

\thanks{Md. Noor-A-Rahim, Fadhil Firyaguna, Jobish John,   Dirk Pesch,  and Holger Claussen  are with the  School of Computer Science \& IT, University College Cork,  Ireland.  (E-mail: {\tt \{m.rahim,ff28,j.john,d.pesch,h.claussen\}@cs.ucc.ie}).

M. O. Khyam is  with the Central Queensland University, Melbourne, Australia.  (E-mail: {\tt m.khyam@cqu.edu.au}).

E. Armstrong is with Johnson \& Johnson, Ireland (E-mail: {\tt earmstr1@its.jnj.com}).

H. V. Poor is with the  Department of Electrical and Computer Engineering, Princeton University, USA (E-mail: {\tt poor@princeton.edu}).

This work was supported  by Science Foundation Ireland under Grant 16/RC/3918 (Confirm Centre for Smart Manufacturing) and Grant 13/RC/2077\_P2 (CONNECT: The Centre for Future Networks \& Communications).
}
}

\maketitle

\begin{abstract}
Industry 5.0 envisions close cooperation between humans and machines requiring ultra-reliable and low latency communications (URLLC). The Intelligent Reflecting Surface (IRS) has the potential to play a crucial role in realizing wireless URLLC for Industry 5.0.  IRS is forecast to be a key enabler of 6G wireless communication networks as it can significantly improve wireless network performance by creating a controllable radio environment. In this paper, we first provide an overview of IRS technology and then conceptualize the potential for IRS implementation in a future smart manufacturing environment to support the emergence of Industry 5.0 with a series of applications. Finally, to stimulate future research in this area, we discuss the strength, open challenges, and opportunities of IRS technology in modern smart manufacturing.
\end{abstract}





\section{Introduction}
Smart manufacturing aims to increase productivity and efficiency by integrating the physical world with the cyber world through the Industrial Internet of Things (IIoT).  The IIoT now connects millions of industrial devices embedded in the physical world to the Internet (or an organization's intranet) and allows for the integration of data generated by them into information systems and business processes and services. This framework is part of the broader trend known as Industry 4.0. The integration of physical and cyber worlds as part of Industry 4.0 is turning traditional industrial automation and control systems into cyber-physical manufacturing systems.

A major driver in the transformation from industrial automation and control into cyber-physical manufacturing systems is the introduction of private 5G wireless networks into industrial environments \cite{Chen_2018}.  Looking further ahead, it is anticipated that the shift from 5G to 6G will also stimulate a transition from Industry 4.0 towards Industry 5.0. While Industry 4.0 will see enhanced introduction of robotics, we consider Industry 5.0 as the next level of human/automation collaboration, where humans and machines share the work, safely and seamlessly, rather than machines replacing humans. In spite of the fact that robots are more reliable than humans and can do more work, they lack many of the fine motor skills that humans have and lack adaptability and critical thinking skills. In the Industry 5.0 era, robots/machines will be used at repetitive, monotonous, dirty, heavy-duty tasks that represent health hazards for humans (e.g., repetitive strain injury, mental health issues, physical injury).  This will free humans up to engage in more stimulating and interesting work, that is harder to automate and requires critical thinking. Thus, Industry 5.0  requires unprecedented collaboration between increasingly powerful and precise machinery and the unique creativity of human beings \cite{Maddikunta2022}. 6G is expected to develop better integration of automatic and high-precision manufacturing processes as well as integrating machines and humans into control loops through low latency and high reliability~\cite{Letaief_2019}.


Compared to traditional wireless communications, industrial wireless communications are already challenged due to metallic structures, electromagnetic interference (e.g., from electrical motor drives or welding apparatus), arbitrary movement of objects (robots and vehicles), room dimension, or thick building structures. On the other hand, full industrial automation requires ultra-reliability and low latency communications in order to deliver sensor data and actuation commands at precise instants with designated reliability (i.e., to perform mission-critical industrial processes). Collaboration of humans and machines in Industry 5.0  will add more complexity to the industrial wireless communication system. In addition,  the rising demand for many emerging services in innovative industries such as AR/VR maintenance, holographic control display systems, etc. will bring forth new communication challenges to  industrial networks \cite{Letaief_2019}. To meet the communication requirements of such services,  Industry 5.0 needs to support advanced technologies such as mmWave/Terahertz communications, advanced localization, and efficient energy harvesting in complex industrial environment.  The key communication system requirements\footnote{In deriving these requirements, we reviewed which applications would most likely be used in future industrial environments and which requirements they will pose on the communication system. These include asset tracking, mobile robots, and condition monitoring, which are all already considered for Industry 4.0 \cite{Chen_2018}, while applications like brain controlled machinery, AR/VR maintenance, immersive collaborative robotics, augmented clothing, and remote haptic interactions are considered for Industry 5.0 \cite{Maddikunta2022}.} for Industry 4.0 and Industry 5.0  are presented in Table \ref{tab:KPI}.

\begin{table*}[h]
	\centering
\caption{Key communications requirements for Industry 4.0 and Industry 5.0.}
\begin{tabular}{||c| c|c||} 
 \hline  \hline
 Key Performance Indicator & Industry 4.0 & Industry 5.0   \\ [0.5ex] 
 \hline\hline
 Data rate & up to 10 Gbps & up to 100 Gbps \\  \hline
Latency & $100$ ms to  $250\mu s$ & less than $100\mu s$ \\  \hline
 Reliability (packet error rate) & $10^{-5}$ to $10^{-8}$ & up to $10^{-10}$\\ \hline
 Connectivity density & 1 device/$m^2$ & 10 devices/$m^2$ \\  \hline
Energy efficiency in communications &  $1\times$ & $10\times$ that of Industry 4.0 \\  \hline
 \hline
\end{tabular} \label{tab:KPI}
\end{table*}

A recently developed concept called intelligent reflecting surfaces (IRSs) can serve as a potential solution to many of the above challenges in future smart manufacturing. An IRS is technology that can significantly improve wireless network performance by creating a programmable radio propagation environment. An IRS is a programmable meta surface containing a large amount of small, low-cost passive antenna arrays that can control a propagating waves' phase, amplitude, frequency, and even polarization. It can increase the efficiency of the wireless network in terms of data rate, coverage, and connectivity. For instance, if a line-of-sight (LOS) link is blocked, an IRS can create a reflective link (i.e., a virtual LOS) to bypass the obstacles between communicating devices. 

This paper aims to conceptualize the potential for IRS implementation in a smart manufacturing environment to support the emergence of Industry 5.0. In Section \ref{sec:SystemModel}, the system model is introduced. Several IRS applications specifically relevant to smart manufacturing are presented in Section \ref{sec:UseCases}. We then outline significant future research directions relating to the challenges and opportunities associated with the use of IRSs in modern smart manufacturing in Section \ref{sec:challenges}, and provide conclusions in Section \ref{sec:Con}.

\section{Intelligent Reflecting Surfaces}\label{sec:SystemModel}

\begin{figure}
\centering
\includegraphics[width=0.95\linewidth]{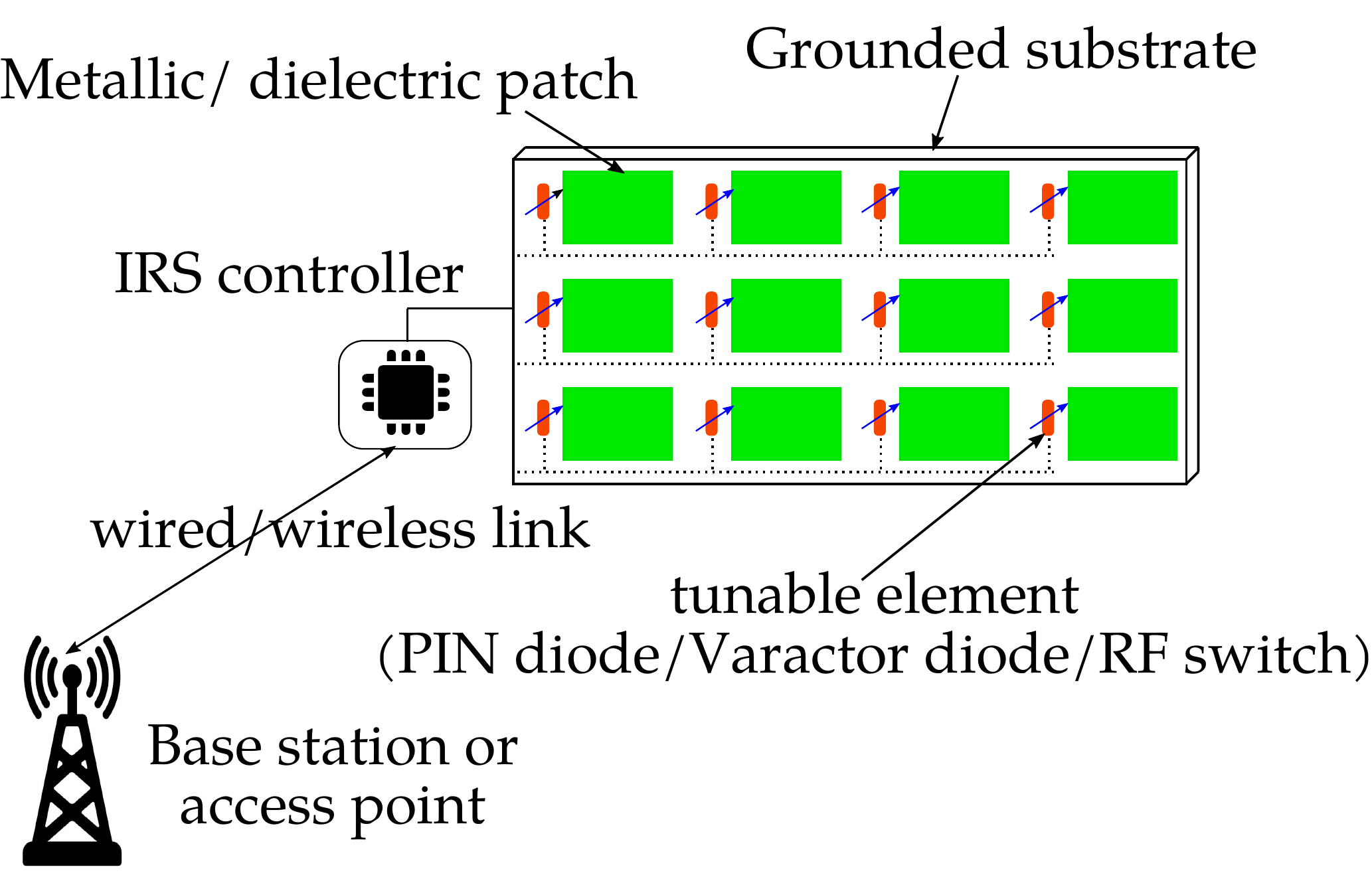}
\caption{IRS Architecture.}
\label{fig:IRS_architect}
\end{figure}


The concept of an IRS is drawn from the concept of a meta surface, which is a 2D form of a meta material. Generally, depending on its structural parameters, this engineered man-made material exhibits unique electro-magnetic properties that cannot be obtained with conventional materials. The IRS is constructed with a large re-configurable array of passive sub-wavelength scale scattering elements (dielectric or metallic patches) that are printed on a grounded dielectric substrate. The size of these patches and their inter-element spacing is usually half of the wavelength or smaller (5 to 10 times smaller) \cite{renzo2020smart}. 

A meta material unit or patch element is capable of adjusting the phase and amplitude of a reflected signal. The direction of reflected signals from each of these elements can be adjusted in the desired fashion (so as to interfere constructively or destructively at the intended location) by controlling their reflection coefficients in real-time. This phenomenon can be characterized by the concept of reflectivity which is defined as the ratio of the reflected signals to the incident signals. The reflectivity of the meta material unit can be obtained by its state, the incident angles and reflected angles.
The reconfigurability of the meta material units or patch elements are achieved with the help of tunable low power electronic circuit elements such as positive-intrinsic-negative (PIN) diodes or varactor diodes or radio-frequency (RF) switches as shown in Fig.~\ref{fig:IRS_architect} \cite{renzo2020smart}. By controlling the bias voltages of the PIN diodes, each PIN diode can be tuned between ON and OFF states which facilitates determining the state of the meta material unit \cite{wu2021intelligent}. In order to program or configure a smart surface or patch elements remotely, an IRS is equipped with a controller (see Fig.~\ref{fig:IRS_architect}). The controller is connected to a base station or access point to receive relevant control and re-configuring commands. Although it is not explicitly  shown in Fig.~\ref{fig:IRS_architect}, an IRS can also be equipped with sensors to help estimate wireless channel conditions \cite{renzo2020smart}. 




Although IRS technology is promising for application in smart industries, it is still immature and there are several open issues that should be addressed including theoretical design and practical integration and engineering manufacture of IRSs. A comprehensive study in this direction can be found in \cite{basar2019wireless}. Also, in Section~\ref{sec:challenges}, we will discuss some open issues associate with the application of IRSs in modern smart manufacturing.

\section{Applications of IRS in smart manufacturing.}\label{sec:UseCases}

In this section we present a number of use cases for an IRS or multiple IRSs in a future manufacturing environment with many autonomous and mobile devices, showing where their functionality can be of particular use.

\subsection{Blockage Mitigation}

Obstructions and blockage are major issues for signal coverage and can cause intermittent and poor connectivity. To circumvent obstructions, an IRS can help steer the incident signals around an obstruction and cover the area shadowed from the base station.
With its large number of passive reflective elements, the IRS enables the adaptation of a wireless environment to overcome the blockage and provide a strong reflective non-line-of-sight (NLOS) link.

\begin{figure}[h]
\centering
\includegraphics[width=0.95\linewidth]{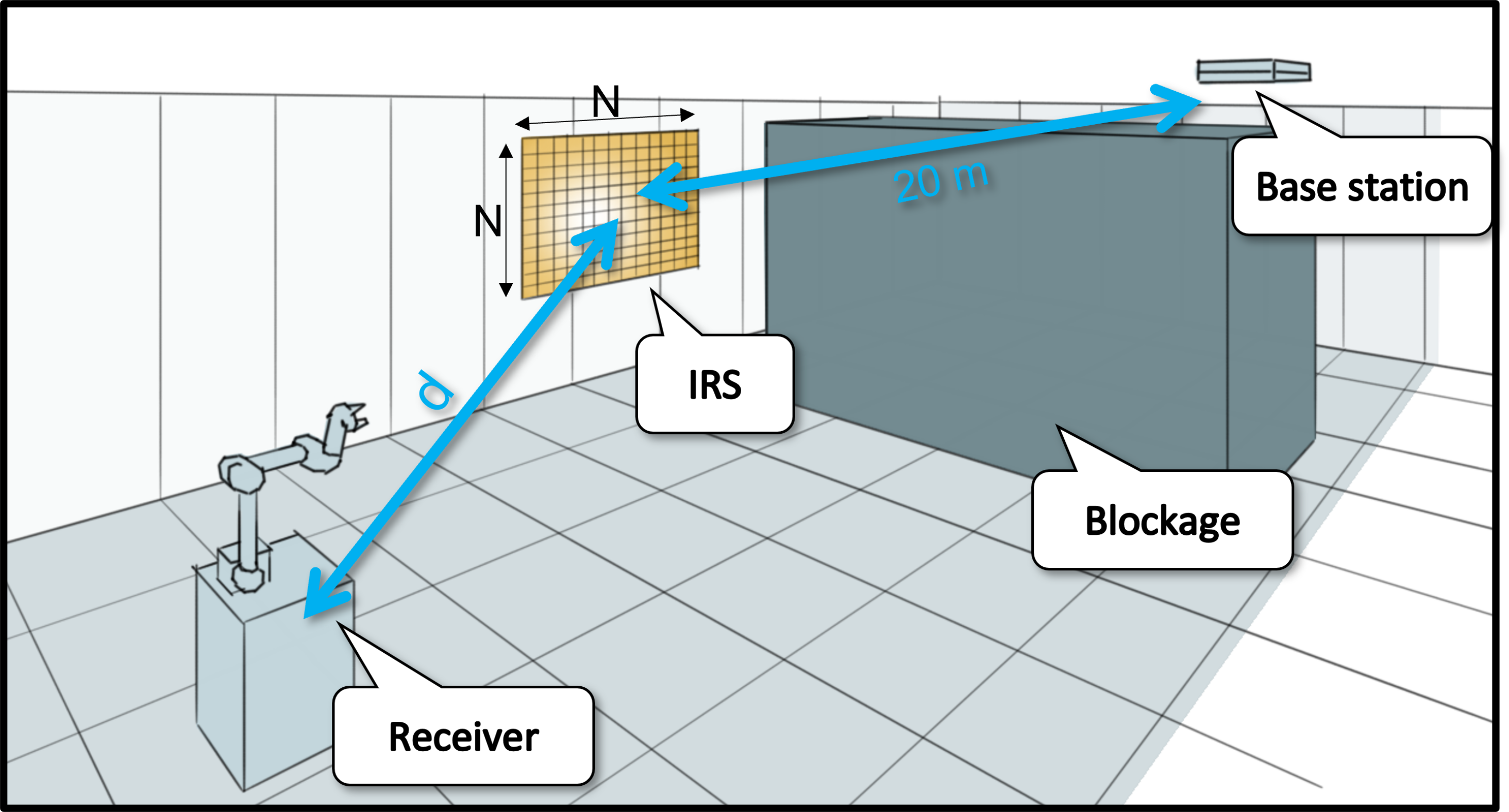}
\caption{Scenario for IRS pathloss analysis.}
\label{fig:IRS_pathloss}
\end{figure}

\begin{figure}[h]
\centering
\includegraphics[width=\linewidth]{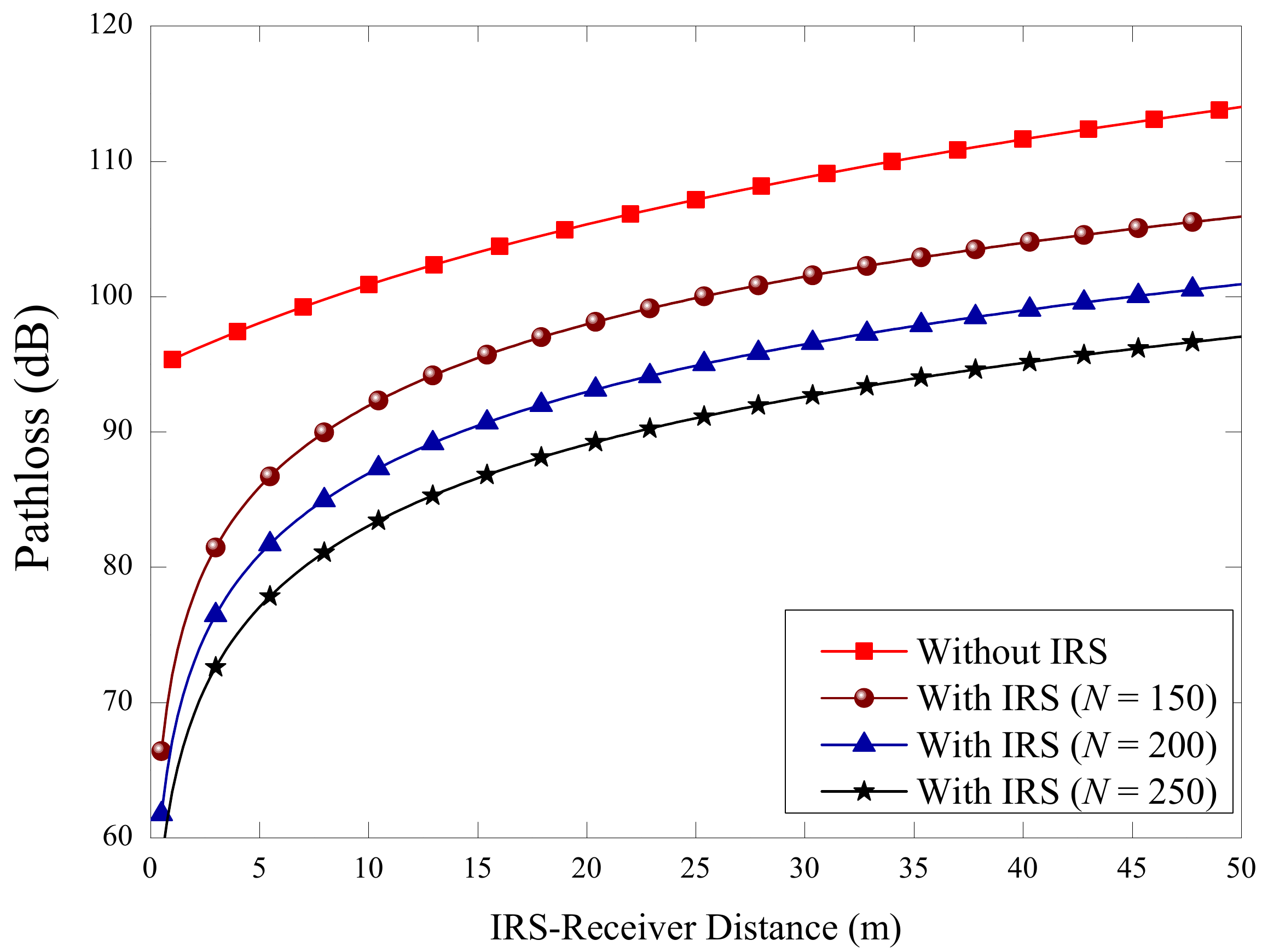}
\caption{Pathloss characteristics of an IRS as a function of the distance $d$ between the IRS and the receiver.}
\label{fig:PL}
\end{figure}

In the following, we show the pathloss characteristics when using an IRS with a transmission frequency of 30 GHz. We assume that the communication link between  base station (BS) and  receiver is completely blocked and an IRS is set up to circumvent the blockage, as depicted in Fig.~\ref{fig:IRS_pathloss}. In our scenario, we assume that the IRS is composed of $N\times N$ elements which are positioned 20m away from the BS in such a way that the IRS has a LOS link with both BS and  receiver.  Fig.~\ref{fig:PL} shows the end-to-end pathloss as a function of the distance $d$ between the IRS and the receiver. This pathloss is calculated using an IRS-based pathloss channel model presented in~\cite{9311936}. We have also plotted pathloss for without IRS scenario, which is calculated using the 3GPP indoor factory NLOS channel model with dense cluttered environment. Significant reduced pathloss is observed with IRS scenarios compared to the without IRS scenario.   With varying numbers of IRS elements, the figure shows that for any given IRS-receiver distance, over 10dB received power gain can be achieved by increasing  $N$ from $150$ to $250$. 

\begin{figure*}
\centering
\includegraphics[width=.9\linewidth]{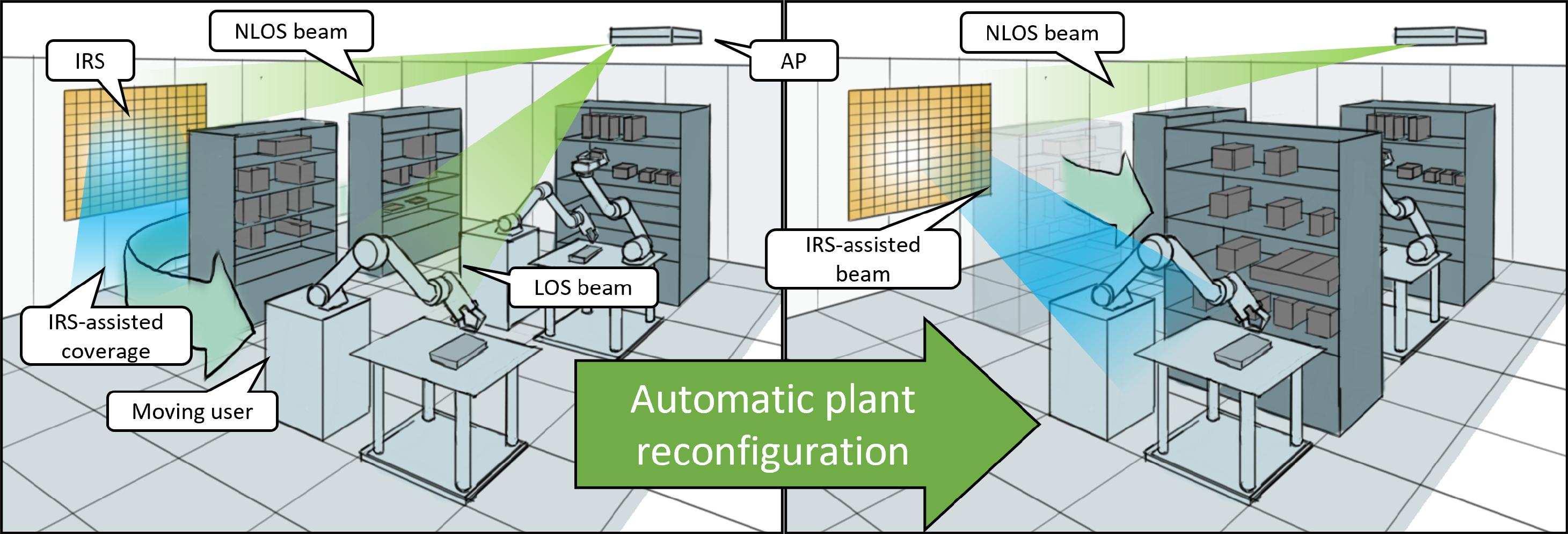}
\caption{Automatic coverage adaptation with an IRS when a factory floor is reconfigured due to production demand.}
\label{fig:blockage_mitigation}
\end{figure*}


This is a positive result to enable flexible smart manufacturing environments, where a plant re-configuration can degrade a given area from being covered by wireless signals to one that is shadowed by machinery.
Hence, the signal coverage can be automatically adapted by steering reflecting beams as the factory plant is reconfigured without the need to redeploy network infrastructure, as illustrated in Fig.~\ref{fig:blockage_mitigation}.
Furthermore, to provide the best radio environment, the IRS could be mounted in a mobile support unit, so its position can be optimized according to the new floor layout.
This flexibility provided by IRSs is key to enable features and applications that rely on high coverage availability described in the following, thus improving communication and operational performance of a smart factory.

\subsection{Millimeter-wave and Terahertz Communications}
Wireless communication for smart manufacturing is characterized by strict link and system requirements regarding 
availability, reliability and latency. For instance, a closed-loop motion control use cases may demand cycle times lower than 1 ms and 99.9999\% service availability for more than 100 nodes \cite{etsi_ts_122104_v1650}.
To support such application requirements, millimeter-wave or terahertz spectrum will provide  wireless channels with wide bandwidth to accommodate large number of nodes operating with high data rate and low latency.
However, propagating mmWave/THz signals suffer from very high path attenuation and lower penetration through materials compared to lower frequency signals in the currently used sub-6GHz bands.
This means that communication links are more vulnerable to blockage, potentially affecting their performance.
Although the use of highly directional antennas can compensate some of the path loss, narrow beam widths may make a link even more vulnerable to blockage.

The mmWave/THz signals have wavelengths at the scale of the surface roughness of many objects, which suggests that scattering may not be neglected as can be at lower frequencies. Also the scattered power relative to the reflected power at mmWave/THz frequencies increases with the incident angle and lower reflection losses (e.g., stronger reflections) are observed as frequencies increase for a given incident angle. 
This means that the signal energy can be more diffused in the environment and the propagation characteristics are highly dependable on surfaces and incident angles of the incident waves.
Here, the application of IRS for mmWave/THz could be a means to generate a more predictable and controllable channel to overcome the effects of scattering. 
To implement this, we need to jointly optimize transmit beamforming and IRS phase shift parameters, maximizing the received power \cite{wang2020intelligent}.

\begin{figure*} [h]
\centering
\includegraphics[width=.7\linewidth]{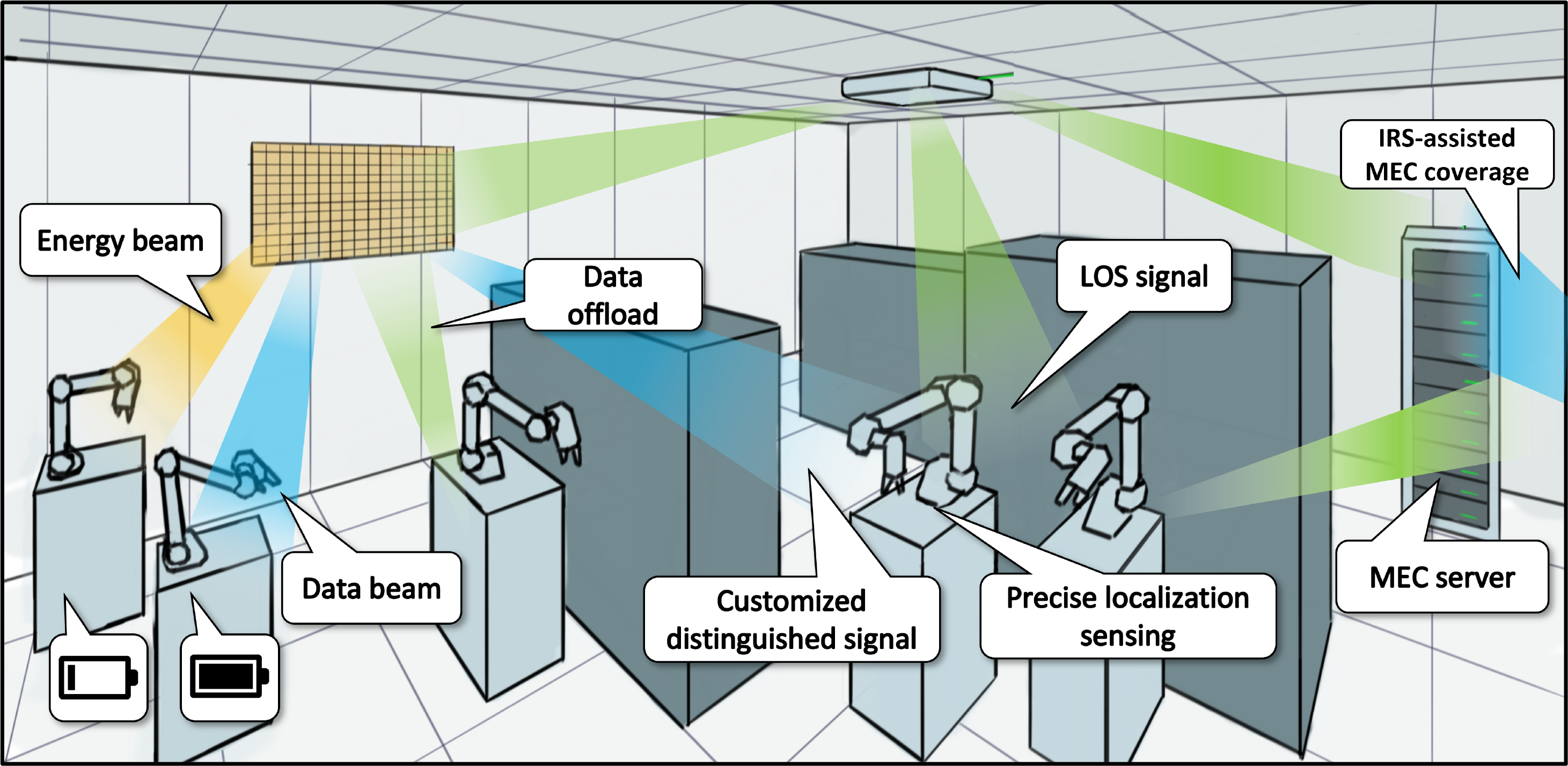}
\caption{IRS-assisted wireless energy transfer, data offloading for mobile edge computing, and high-accuracy localization.}
\label{fig:IRS_energy_localization_mec}
\end{figure*}%

\subsection{Wireless Energy Transfer}
The envisioned future smart manufacturing environment will consist of many wireless sensors and actuators, which traditionally are powered by batteries, which deplete over time. Replacing depleted batteries of thousands of sensors and actuators is a costly task. Additionally, many sensors may be installed in sensitive places that are not suitable for frequent battery replacements. The use of wireless energy harvesting, which allows sensors/actuators to harvest power from a signal that was intended for data transmission or power transmission, has shown to be an excellent solution to address this issue. The problem with wireless power transfer, however, is the drop in wireless power over large transmitter to receiver distances. A wide range of techniques have been proposed to overcome this power loss, such as waveform design, energy transmission and scheduling, and energy beamforming, which can be implemented at the transmitter and/or receiver in order to improve the efficiency of wireless power transfer.  Unfortunately, many of the above solutions are not ideal for smart manufacturing environments, as they require significant signal processing capabilities, which many low power IIoT devices do not possess. 

Smart manufacturing could benefit from wireless power transfer enhanced by IRSs. Thanks to the deployment of IRSs close to IIoT devices, the issue of high path loss can be effectively alleviated by creating an energy-efficient charging zone for those devices, as depicted on the left side of Fig.~\ref{fig:IRS_energy_localization_mec}.  When IRSs are deployed correctly in LOS with transmitters and receivers and their beamforming capabilities are fully exploited, the received power of nearby IIoT devices can be substantially increased. An energy receiver can take advantage of an IRS's passive beamforming to improve the transmission efficiency of wireless energy while simultaneously enhancing the signal strength at an information receiver. Moreover, it realizes the possibility of improving both the rate and energy performance in wireless power transfer by increasing the wireless charging efficiency. This in turn  helps to reduce the transmit power and provides more flexibility in the design of transmit beamforming for information receivers. The effectiveness of passive beamformers for wireless power transfer is expected to be crucial in practice.  To achieve their benefits, however, they require channel state information at the energy transmitter.

\subsection{Sensing \& Localization}
Sensing and localization in smart manufacturing creates the opportunity to continuously monitor individual products, providing the possibility for product customisation through tighter control, management and analysis of critical manufacturing parameters. Thus, acquiring the precise location of objects and being able to sense local information and ambient parameters in the environment in industrial settings is becoming indispensable to enable location and sensing-based services and applications. For example, the transparent production and logistics processes of  smart manufacturing can be improved significantly by knowing what is happening when, where, and how. Automated guided vehicles can improve production supply, assembly lines (through transport platforms) and warehouse logistics systems. IRSs offer significant advantages for precise localization (e.g., using angle of arrival method) and high-resolution sensing solutions in industrial settings since it can actively customize the propagation environment, as illustrated in right side of Fig.~\ref{fig:IRS_energy_localization_mec}. The underlying idea of wireless signal sensing is based on the principle that receivers can identify the effects that sensing targets have on  wireless signal propagation. The receiver exploits the observations to detect target behaviour. Unlike conventional sensing techniques, IRS-assisted sensing creates a controllable radio environment in preferred directions interacting with sensing targets. As a result, IRS-assisted sensing does not require a LOS link between the receiver and the sensed target~\cite{Hu_2020}.  On the other hand, in IRS-assisted localization, an IRS is deployed between the access point (AP) and receiver in such a way that the AP can investigate a user's reflected signal through various IRS configurations to achieve accurate locations.




\subsection{Mobile Edge Computing}

IRSs can support mobile edge computing (MEC) in smart manufacturing. The MEC paradigm extends computing resources from the cloud to the network's edge. Future smart factories will have a very large number of wireless devices generating large volumes of data in real-time. Generally, these devices do not possess the required processing power and battery capacity, creating the need to offload processing operations to the edge servers (preferred) or cloud platforms. This helps to reduce the end to end delay and avoids unwanted network congestion~\cite{mec_Zheng}. 

An IRS can help establish strong wireless computation offloading links, reducing packet losses/re-transmissions and enhancing spectral and energy efficiency. As shown in Fig.~\ref{fig:IRS_energy_localization_mec}, an IRS-assisted MEC system in a smart factory consists of one or more access points/base stations with co-located edge computing nodes/servers, IRS elements along with its controller, and large numbers of field devices. MEC servers are usually co-located with the access points for ease of joint optimization of computational and communication resources. Usually, in an industrial environment, the access points mounted on the ceiling are connected to the edge servers on the shop floor over a wired connection (e.g. fiber). This limits the flexibility of the smart factory. The application of IRSs, along with upcoming wireless technologies (5G/6G), helps replace this with a completely wireless link with ultra-high reliability, low latency and high bandwidth. There can be several distributed edge computing servers on the shop floor, and the wireless devices can offload their computations to their nearby servers wirelessly. A virtual LOS link with enhanced channel gain can be established between the field devices and the AP/edge server by adequately tuning the IRS reflecting elements, which helps to offload data to the MEC server more quickly. The processed results/control actions are also quickly fed back to end nodes, shortening the end-to-end delay. Currently, in many smart manufacturing applications, local node processing is adopted due to weak communication links resulting in idle resources at the edge. Thus IRSs can help to exploit powerful edge computational resources better by making them more accessible to an increased number of wireless field devices. However, carrying out the optimal computational resource allocation at the edge servers along with the communication resource allocations at the BS/AP and the dynamic tuning of IRS reflection coefficients will be a major challenge.

\section{Challenges \& Open Issues} \label{sec:challenges}

\subsection{Environment-aware Passive Beamforming}
One of the main challenges for the successful application of IRSs in smart manufacturing is designing environment-aware and dynamic passive beamforming. 
In practice the design of IRS passive beamforming is determined by the discrete amplitude and phase-shift levels of each element. A beam steering process requires coordination of the phase control of individual scattering elements. Despite  limited phase shifts available at an individual IRS scattering element, an IRS with a large number of scattering elements can enable more flexible phase tuning. However, computational complexity is a price to pay for such flexibility. Moreover, a larger number of scattering elements means greater difficulty in channel estimation, which could hinder efficient phase control.  

While exhaustive search may provide the best solution for determining the best amplitude/phase-shift levels, the approach is computationally complex, and may be infeasible for scenarios where energy savings are paramount. Efficient algorithms are therefore required to  estimate channels and control phase shifts of all scattering elements in real-time following the dynamics of the radio environment.  A practical solution as opposed to exhaustive search can be achieved by solving the problem with continuous amplitude and phase-shift values, and then calculating the closest discrete values of the obtained solutions \cite{Wu_2020}.

To realize practical and efficient IRS beamforming, machine learning approaches can assist to effectively resolve the above problems by using locally observed information in the smart manufacturing environment. High numbers of scattering elements and sensors means that a significant amount of information can be collected during channel sensing, facilitating machine learning approaches based on large data sets. The use of data driven machine learning has the potential to   minimize   the overhead of information exchange between the IRS and active transceivers.  In an IRS-based smart manufacturing environment, however, machine learning approaches must be designed to fit the hardware constraints. Large-scale experimental evaluations are essential to gain more insights into the effectiveness of passive beamforming of IRS in real-world deployments.


\subsection{Radio Resource Management}
In any wireless network, one of the most important tasks is to allocate  radio resources optimally. In general, IRS radio resource management is mainly concerned with power allocation, bandwidth allocation, and node-IRS connectivity.
Due to the specific dynamics of interference in IRS-enabled wireless environments, transmit power allocation is an essential component for the effective operation of an IRS enabled environment. In smart manufacturing, numerous wireless devices embedded in machines, autonomous vehicles, and the environment coexist in close proximity, making this an even more pressing problem. In order to minimize interference while maximizing the system's capacity,  effective power allocation approaches need to be developed. On the other hand, bandwidth allocation determines the most suitable allocation of users to different sub-channels to increase bandwidth efficiency. Due to the frequency-agnostic nature of IRS elements, one common IRS reflection matrix is shared among sub-channels, making optimization problematic. In order to address the problem, dynamic passive beamforming can be used. In this scheme, the resource blocks are dynamically assigned to different user groups with different IRS phase shifts for different time slots.


Another major challenge in enhancing the wireless network performance using IRSs lies in associating users/wireless devices to IRSs and selecting their communication mode. Some wireless devices may have an excellent direct LOS link with the BS and hence need not be associated with any IRS. Other devices may make use of single reflection links for better network performance and need to be associated with either the user-side or BS-side IRS (detailed in Section~\ref{label_deployment}), while again others may take advantage of both the single reflection and double reflection links. Most of these wireless devices in a smart factory are highly mobile, which results in a highly challenging dynamic IRS-user association. To assign users optimally to different IRSs, the channel state information (CSI) of all communication links is essential, which is very difficult to obtain in practice. How beneficial it is to integrate sensing devices onto IRSs for channel sensing, making it semi-passive, is another question that needs to be investigated. Another challenge lies in establishing a reliable wireless communication link between the IRS controller and the BS. Industry 5.0 forecasts the replacement of the rigid wired communication links in a smart factory, questioning use of wired backhaul links between IRSs and BSs, especially when they are distributed across the shop floor.


\subsection{Channel Characterization}
There are two major challenges for end-to-end analysis of an IRS system. Firstly, to analyze the performance limits of an IRS link, new channel propagation models are needed to obtain the link budget analysis. Path loss models depend on several parameters including the size of the IRS and the mutual distances between the transmitter/receiver and the IRS \cite{tang2020wireless}. 

Secondly, to decode the signal reflected by the IRS, the channel should be properly estimated. In addition to estimating the direct link between transmitter and receiver, two IRS-assisted channels need to be estimated, i.e., the transmitter-IRS and IRS-receiver channels, and they cannot be separately estimated via traditional training-based approaches in general because IRSs are typically passive and cannot perform channel estimation by themselves. As a result, alternative approaches are needed to perform channel estimation, while keeping complexity and overhead of IRS operations as low as possible \cite{renzo2020smart}. The problem becomes even more challenging with large IRS arrays since the time overhead to perform the channel estimation increases linearly with the number of IRS elements \cite{wang2020channel}. Furthermore, such channel models and estimation approaches should consider the specific industrial environment. In a factory floor environment, the presence of metallic surfaces on machinery furniture and vehicles leads to a wide range of values for channel parameters, such as path loss and multipath parameters, specially when using high frequency signals. The applicability of the current/new models (both electromagnetic material models and the IRS assisted wireless channel models) needs to be validated for the manufacturing environment. Future research should also concentrate on efficient environment aware dynamic channel characterization approaches. 



\subsection{Deployment Issues}
\label{label_deployment}
IRSs can be deployed in an industrial environment using different strategies; (i) close to the distributed wireless devices (known as user-side IRS deployment), (ii) close to the base station (known as BS-side IRS deployment), or (iii) in a hybrid style combining both the user-side and BS-side IRS deployment~\cite{you2020deploy}. Each has its pros and cons. User-side IRS deployment provides enhanced network coverage mainly for the users or wireless devices within its local vicinity. In contrast, the BS-side deployment provides extended network coverage. One of the main motivations for using an IRS is to provide a virtual LOS link between base station and wireless devices whenever obstacles are present. The placement of a user-side IRS is relatively easy to establish a virtual LOS link between BS and intended local users, whereas placing a BS-side IRS to establish a virtual LOS link for all its users is difficult. The communication signalling overhead between the IRS controller and the BS for tuning the reflection coefficients of the IRS is relatively low for BS-side deployment due to their close proximity. Hybrid IRS deployment combines the advantages of both user-side and BS-side schemes. It also helps to exploit double reflection links (inter-IRS reflection links) to provide more LOS paths between the served users/wireless devices in a smart factory and the BS/AP. At the same time, a hybrid deployment scheme brings additional complexity in the design, deployment and management of an IRS. However, the main challenge with these three options is exactly where and how to deploy them. IRSs may be deployed in a centralized or distributed fashion (for a given number of reflecting elements), and it is not yet clear which approach is the best for an industrial environment.  An IRS’s low cost provides the flexibility to opt for a dense deployment on a factory floor if required. However, their joint network performance optimization will be a challenging task.

\section{Conclusion}\label{sec:Con}

In this article, we have discussed the prospects of IRS-aided wireless networks in a smart manufacturing environment to support the evolution towards Industry 5.0 by unfolding their potential features and advantages through different wireless network scenarios. As IRS technology is still in its infancy, we have elaborated on the most pressing challenges as well as the potential opportunities for research into future IRS-aided wireless factory automation. Thus, it is hoped that this article will serve as a useful and inspiring resource for future research on IRS-based smart manufacturing to unlock its full potential in a future industrial environment.

\bibliography{references}
\bibliographystyle{IEEEtran}

\section{Biography}

\textbf{Md. Noor-A-Rahim} received the Ph.D. degree from the Institute for Telecommunications Research, University of South Australia, Australia in 2015. He is currently a Research Fellow with the School of Computer Science \& IT, University College Cork, Ireland. 

\textbf{Fadhil Firyaguna} received his Ph.D. degree from the Trinity College Dublin, Ireland in 2020. He is a post-doctoral  researcher with the School of Computer Science \& IT, University College Cork, Ireland.


\textbf{Jobish John} received his Ph.D in Electrical Engineering from the Indian Institute of Technology, Bombay in 2020. He is currently working as a researcher with the School of Computer Science \& IT, University College Cork, Ireland.


\textbf{Mohammad Omar Khyam} received the Ph.D. degree from the University of New South Wales, Australia in 2015. He is currently with the Central Queensland University, Melbourne, Australia.

\textbf{Dirk Pesch} is Professor in the School of Computer Science \& IT at University College Cork. He holds a Dipl.Ing. degree from RWTH Aachen University, Germany, and a PhD from the University of Strathclyde, Glasgow, Scotland.


\textbf{Eddie Armstrong} is an Engineering Fellow with 
Johnson \& Johnson. He received his Ph.D. degree from the Computer Science and Information Systems department, University of Limerick, Ireland. 

\textbf{Holger Claussen} is Head of the Wireless Communications Laboratory at Tyndall National Institute and a Research Professor in the School of Computer Science \& IT at University College Cork, Ireland. 

\textbf{H. Vincent Poor} is the Michael Henry Strater University Professor of Electrical Engineering with the Faculty at Princeton University. He received the Ph.D. degree in electrical engineering and computer science from Princeton University, Princeton, NJ, USA, in 1977.

\end{document}